\documentclass[journal=jacs,manuscript=article]{achemso}

\usepackage{xr}
\usepackage[T1]{fontenc}

\usepackage{subcaption}
\usepackage{graphicx}
\usepackage{amsmath}
\SectionNumbersOn

\usepackage{lscape}

\newcommand{ \mrm }[1]{\mathrm{#1}}

\newcommand{ \etal }{\textit{et al.}}

\newcommand{ \mgo   }{MgO}
\newcommand{ \mgoh  }{Mg(OH)$_2$}
\newcommand{ \mgion }{Mg$^{2+}$}
\newcommand{ \SI    }{Supporting Information}

\title{Comprehensive Molecular-level Understanding of MgO Hydration through Computational Chemistry}

\author{Taichi Inagaki}
\email{taichi.inagaki@keio.jp}
\phone{+81 (0)45 566 1710}
\affiliation{Department of Chemistry, Faculty of Science and Technology, 
Keio University, 3-14-1 Hiyoshi, Kanagawa 223-8522, Japan}

\author{Miho Hatanaka}
\affiliation{Department of Chemistry, Faculty of Science and Technology,
Keio University, 3-14-1 Hiyoshi, Kanagawa 223-8522, Japan}
\alsoaffiliation{Institute for Molecular Science, Okazaki, Aichi 444-8585, Japan}

\begin{document}

\clearpage

\begin{abstract}  
 The hydration of magnesium oxide (MgO) to magnesium hydroxide (\mgoh) is a fundamental 
 solid-surface chemical reaction with significant implications for materials science.
 Yet its molecular-level mechanism from water adsorption to \mgoh{} nucleation and growth 
 remains elusive due to its complex and multi-step nature.
 Here, we elucidate the molecular process of MgO hydration based on structures 
 of the MgO/water interface obtained by a combined computational chemistry approach of 
 potential-scaling molecular dynamics simulations and first-principles calculations 
 without any \textit{a priori} assumptions about reaction pathways.
 The result shows that the \mgion{} dissolution follows the dissociative water
 adsorption.
 We find that this initial dissolution can proceed exothermically even from 
 the defect-free surface with an average activation barrier of $\sim$12 kcal/mol.
 This exothermicity depends crucially on the stabilization of the resulting 
 surface vacancy, achieved by proton adsorption onto neighboring surface 
 oxygen atoms. 
 Further \mgion{} dissolution then occurs in correlation with proton penetration 
 into the solid.
 Moreover, we find that the \mgoh{} nucleation and growth proceeds according to 
 the dissolution-precipitation mechanism, rather than a solid-state reaction mechanism
 involving a direct topotactic transformation.
 In this process, \mgion{} ions migrate away from the surface and form amorphous 
 Mg-OH chains as precursors for \mgoh{} nucleation. 
 We also demonstrate that sufficient water facilitates the formation of more 
 ordered crystalline nuclei.
 This computational study provides a comprehensive molecular-level understanding of 
 MgO hydration, representing a foundational step toward elucidating the mechanisms 
 of this class of complex and multi-step solid-surface chemical reactions.
\end{abstract}

\clearpage
%
%
\section{Introduction}  \label{sec:Intro}
Chemical reactions on solid surfaces are fundamental processes and often involve 
structural transformations of the solid material itself. 
Such transformative reactions are ubiquitous, ranging from natural phenomena such as 
metal corrosion and mineral weathering to industrial applications such as heterogeneous 
catalytic reactions.
While these reactions can be understood from a macroscopic perspective 
as a simple phase change based on thermodynamic chemical equilibrium
\cite{Giauque1937,Chen2002,Chang2004}, 
at the microscopic level they involve many elementary processes 
(e.g., chemical bond breaking/forming and atomic migration) in a multi-step and 
non-equilibrium manner over a wide range of time scales. 
Due to this complexity, understanding such reactions at the molecular level 
remains a formidable challenge for both experimental and theoretical/computational 
approaches.
Their detailed elucidation is therefore an important objective in interfacial chemistry.

One such chemical reaction is the hydration of magnesium oxide (\mgo) 
to magnesium hydroxide (\mgoh), MgO + H$_2$O $\rightarrow$ \mgoh. 
This exothermic transformation of rock-salt MgO into \mgoh{} nanosheets 
($\Delta H \sim -81$ kJ/mol) involves a simply structured solid and ubiquitous water.
Owing to this structural and electronic simplicity, this reaction has been used as a model 
system to study the interfacial properties of oxide-water systems\cite{water_at_interface}. 
In addition, this hydration reaction is deeply related to various industrial technologies, 
including heat storage\cite{Yan2015,Alfonso2019,Kato1996}, cement production
\cite{Lea1998,Dung2018,Ma2020}, and catalysis\cite{Cadigan2013,DiCosimo2014}, 
and thus also holds practical significance. 
However, the detailed mechanism at the molecular level remains elusive,
hindering efforts to precisely control the reaction's thermodynamics and kinetics 
for industrial purposes
\cite{Kato1996,Kondo2021,Pettauer2024}.

Early studies of the reaction between MgO surfaces and water have primarily focused on 
the interfacial structure of water and its molecular dissociation. 
This hydration reaction starts with the adsorption of water molecules on the surface. 
The adsorption structure of water molecules at low temperatures has been investigated 
in detail in previous studies\cite{Wu1992,Xu1997,Liu1998,Ferry1998,Finocchi2008,Giordano1998,
Odelius1999,Site2000,Jug2007,Radoslaw2011}, and the structure has been almost established. 
In particular, the well-ordered water monolayer structures with $c$(4$\times$2) and 
$p$(3$\times$2) symmetries have been revealed by theoretical and experimental studies
\cite{Radoslaw2011}. 
Molecular-level analyses have also shown that water molecules can easily dissociate, 
even in the monolayer on the perfect MgO (100) surface, by accepting a strong hydrogen bond 
(HB) from a neighboring water molecule\cite{Odelius1999}. 
In recent years, a stable adsorption structure under ambient conditions was proposed to be 
$p$(3$\times$2) symmetry with eight water molecules involving 25\% dissociation\cite{Sassi2024}, 
and the structure was compared with the vibrational sum frequency generation spectroscopy 
measurements\cite{Adhikari2021}. 
These extensive studies have greatly advanced the understanding of the adsorption and 
dissociation processes of water molecules. 

While the initial dissociative water adsorption under ambient conditions
are becoming well-understood,
the subsequent elementary processes on the MgO/water interface remain less clear. 
A crucial next step is widely considered to be the dissolution of \mgion{} ions 
from the surface. 
Following On$\mrm{\check{c}\acute{a}}$k and co-workers\cite{Oncak2016},
we use "dissolution" also in the sense that \mgion{} ions leave the surface layer 
and are located above the MgO surface.
At the molecular level, theoretical studies found locally stable reconstructed 
interfaces with \mgion{} ions above the surface\cite{Mejias1999,Jug2007_MD,Jug2007_path,
Oncak2015,Oncak2016,Ishida2024}, which appears relatively consistent 
with the experimental findings\cite{Carrasco2010}. 
In particular, these calculations have identified various intermediate states, 
such as semi-detached \mgion{} ions at the low-coordinated edge sites\cite{Mejias1999} or 
hydrated/hydroxylated contact ion pairs above the surface\cite{Oncak2015}, 
as likely exothermic precursors to dissolution and \mgoh{} formation. 
These observations suggest that the dissolution of \mgion{} ions plays an important role 
in the early stage of the hydration. 
However, a significant limitation in these studies is that the dissolution is induced 
artificially within the theoretical calculations. 
Therefore, it remains inconclusive whether \mgion{} dissolution is the definitive 
spontaneous process following water adsorption and what the key factors for exothermic 
dissolution are. 

Another important focus for further understanding the hydration is the nucleation 
and growth mechanism of \mgoh{} crystals, which is still a controversial issue
\cite{Wogelius1995,Liu1998,Oviedo1998,Mejias1999,Lee2003,Rocha2004,Jug2007_path,Sasahara2015,
Kuleci2016,Ishida2024,Amaral2010,Kondo2021,Chen2017,Tang2018,Maltseva2019,Bracco2024}. 
One of the notable mechanisms is a direct transformation of the solid structure 
based on a topotactic relationship between the MgO (111) and \mgoh{} (0001) planes
\cite{Feitknecht1967}. 
In this study, we refer to this as the solid-state reaction mechanism. 
This mechanism posits that the (0001) planes in \mgoh{} are formed 
parallel to the (111) planes in MgO.
Wogelius \etal{} proposed\cite{Wogelius1995}, 
on the basis of the results of their elastic recoil detection analysis and theoretical 
calculations, that the hydration proceeds by the diffusion of bulk \mgion{} ions 
outward and protons (H$^+$) from water molecules inward in the direction parallel 
to the MgO (111) planes. 
The diffusion processes were also expected to be realized together with the creation 
of a penetrative hydroxylated channel with help from pits\cite{Liu1998}, 
and the channels would result in the interlayer spaces in \mgoh{} crystals. 
This mechanism has also been examined in theoretical studies\cite{Refson1995,Oviedo1998, 
Mejias1999,Asaduzzaman2020}.
For instance, some studies have mentioned that parts of the reconstructed MgO surface 
structures closely resemble the \mgoh{} structure\cite{Refson1995,Oncak2016,Asaduzzaman2020}.
In particular, Jug and co-workers presented an atomistic process where the MgO surface 
is dug along the (111) plane by diffusion of \mgion{} ions and protons and eventually 
the (0001) surface of \mgoh{} is formed along the MgO (111) plane\cite{Jug2007_path}. 
Recently, Ishida and Ishimura revisited the hydration process proposed 
by Jug \etal{} using more accurate (density functional theory) calculations\cite{Ishida2024}.
In their study, they suggested that the rate-determining step in this mechanism is 
an extraction of Mg atoms from an inner layer to the aqueous surface layer.
It should be noted that these calculations are again based on an artificially 
determined reaction pathway and therefore are not necessarily examples of the formation 
of crystal nuclei.
In addition, they have not demonstrated the formation of well-ordered \mgoh{} nuclei.

An alternative explanation is that nucleation and growth proceeds according to 
a dissolution-precipitation mechanism.
In this process, a hydroxylated MgO surface layer dissolves in water,
leading to the aggregation of dissolved species and subsequent nucleation.
These nuclei then precipitate due to the low solubility of \mgoh.
This dissolution-precipitation mechanism was proposed at an early time\cite{Vermilyea1969},
and the relevant insights are still being provided by experimental studies
\cite{Amaral2010,Baumann2015,Kondo2021,Chen2017,Tang2018,Maltseva2019,
Luong2023_nanoscale,Luong2023_Langmuir,Bracco2024}. 
For instance, Luong and co-workers demonstrated that a thick water layer on the solid 
surface is crucial for facilitating the reactions\cite{Luong2023_nanoscale,Luong2023_Langmuir},
which indicates that limited water availability delays nucleation.
In addition, Bracco and co-workers showed that while a reaction layer forms rapidly
on the MgO surface even under relatively dry conditions, 
it does not continue to grow over time\cite{Bracco2024}. 
They also indicated that the layer on the surface consists of 
amorphous or poorly crystalline phases rather than crystalline \mgoh\cite{Bracco2024}. 
Their observations imply that the nucleation and growth is favored in an aqueous environment.
Overall, recent experimental studies seems to adopt the dissolution-precipitation mechanism, 
without examining the possibility of the solid-state reaction mechanism, 
as a foundational premise for their analysis.
An investigation by Chen \etal{} is one of the few theoretical studies
\cite{Chen2017,Luong2023_Langmuir} to follow the dissolution-precipitation mechanism. 
Although their calculations focused only on small clusters, they suggested that 
the self-assembly of extremely small sheet-like \mgoh{} clusters is a possible pathway 
for the formation of larger \mgoh{} nanoparticles\cite{Chen2017}. 
Thus, while various studies have focused on these different mechanisms, 
elucidating which mechanism is more plausible through direct comparison remains 
a pivotal and unresolved question for achieving a molecular-level understanding 
of MgO hydration.

In this study, we investigate molecular processes in MgO hydration based on 
computationally obtained MgO/water interfacial structures. 
We consider an MgO solid surface with the (100) plane initially covered 
by a water film of two-monolayer thickness. 
The (100) plane of MgO is known to be the most stable 
surface in the atmospheric environment. 
According to previous experimental and theoretical studies
\cite{Ewing2006,Luong2023_Langmuir,Oncak2016}, 
a water coverage of two monolayers corresponds to a relatively high humidity environment 
(relative humidity > 70\%) and is probably the minimum to observe \mgion{} 
dissolution and the subsequent hydration processes. 
This model system is also motivated by a heat storage technology using the MgO hydration 
reaction\cite{Yan2015,Alfonso2019,Kato1996,Ishida2024}. 
The previous computational studies of the hydration\cite{Jug2007_path,Oncak2015,
Asaduzzaman2020,Ishida2024} have relied on pre-assumed reaction pathways, 
leaving the molecular mechanisms of this slow 
(hours to days\cite{Kato1996,Mejias1999,Lee2003}) process unclear. 
In this work, to go beyond the previous works and elucidate the reaction mechanism,
we use an advanced molecular simulation method, potential-scaling molecular dynamics 
(PS-MD), and first-principles calculations, and examine locally stable structures 
along the hydration without any $\textit{a priori}$ assumptions.
This paper begins with an overview of the interfacial structural changes 
along the hydration. 
We then investigate the molecular details by classifying the observed structures 
into three different stages: (1) water adsorption on the MgO surface, 
(2) \mgion{} dissolution into the water layers, and (3) \mgoh{} nucleation. 
In the first stage, we examine the interfacial water structure,
including its HB networks and molecular orientations, to elucidate the mechanism
of water dissociation under ambient conditions.
In the second stage, mainly representing the initial \mgion{} dissolution, 
we analyze how this process occurs and identify the origins of its exothermicity 
by calculating the corresponding energy profiles.
Finally, by investigating the third stage, 
we provide molecular insights into the nucleation and growth of \mgoh. 
Our results suggest that the dissolution-precipitation mechanism 
is favored over the solid-state reaction mechanism in this system. 
To the best of our knowledge, this is the first computational study showing that 
the well-ordered \mgoh{} nanosheets can form from an amorphous precursor phase 
in an aqueous solution. 

%
%
\section{Computational Methods}\label{sec:Method}
This section briefly describes the computational methods used in this study.
Full details regarding the methods and validations are provided in \SI{}.

We first generated various structures relevant to the hydration of MgO using 
the PS-MD simulation method. 
This simulation method allows us to induce various structural changes, including 
barrier-crossing processes of chemical bond breaking/forming and atomic migration, 
by repeating short MD simulations with continuously scaling the potential energy 
surface\cite{Inagaki2022}. 
The simulation consists of a repetitive cycle in which the potential energy 
surface is gradually flattened and then restored. 
As detailed in Section \ref{sec:Results}, this technique captured key elementary 
processes, such as the dissolution of \mgion{} ions from the surface and 
the penetration of protons into the bulk. 
The observation of these events demonstrates that our simulation accesses timescales 
inaccessible to conventional MD simulations.
The present PS-MD simulations were performed using ReaxFF\cite{Duin2001_ReaxFF}, 
a low-cost classical force field that can handle bond recombination.
To generate interfacial structures along the hydration, we performed a total of 1,000 
PS-MD cycles. 
From the trajectory of each cycle, we extract the last snapshot, 
yielding a total of 1,000 structures. 
These 1,000 structures were then fully optimized using the same ReaxFF potential 
to identify locally stable minima.

After the PS-MD simulations and the subsequent optimization calculations 
at the ReaxFF level, the structures were improved using the density functional 
based tight binding (DFTB) method\cite{Elstner1998}. 
The DFTB method provides a semi-empirical electronic structure 
calculation that is two to three orders of magnitude faster than density functional 
theory (DFT) calculations. 
Furthermore, for all structures that were more stable than the initial structure 
in the DFTB calculations, a further refinement 
using the more accurate DFT method was performed to select the structures 
to be analyzed for the hydration. 
We adopted the BLYP exchange-correlation functional, which has been very widely used 
in various researches from materials science to life science and is known to perform 
well in hydrogen-bonded systems\cite{Sprik1996,Odelius1999,Site2000}. 

The simulation system was the same for the three calculation levels (ReaxFF, DFTB, and DFT). 
The solid surface was modeled as a five-layer MgO slab with the perfect (100) surface. 
In order to investigate the hydration reaction, a water film of two-monolayer thickness 
was placed on the MgO slab surface. 
The size of the slab in the lateral $xy$ plane was 16.85 \AA{} $\times$ 16.85 \AA{} based 
on the experimental lattice constant of MgO (4.21 \AA\cite{Hazen1976_MgOlattice}), 
and the thickness of the slab, including the water film, was about 17 \AA. 
Using this simulation system, we performed five independent runs, each consisting of 
a PS-MD simulation followed by geometry optimizations at the ReaxFF, DFTB, and DFT levels. 
In the next section, we will investigate the molecular mechanism of the hydration
on the basis of the MgO/water interfacial structures finally obtained by the DFT calculations.

%
%
\section{Results and Discussion} \label{sec:Results}
\subsection{Overview of the MgO hydration process} \label{sec:results_overview}
\begin{figure}
  \begin{center}
    \includegraphics[width=1.0\linewidth,trim=2.5cm 0cm 3cm 0cm, clip]{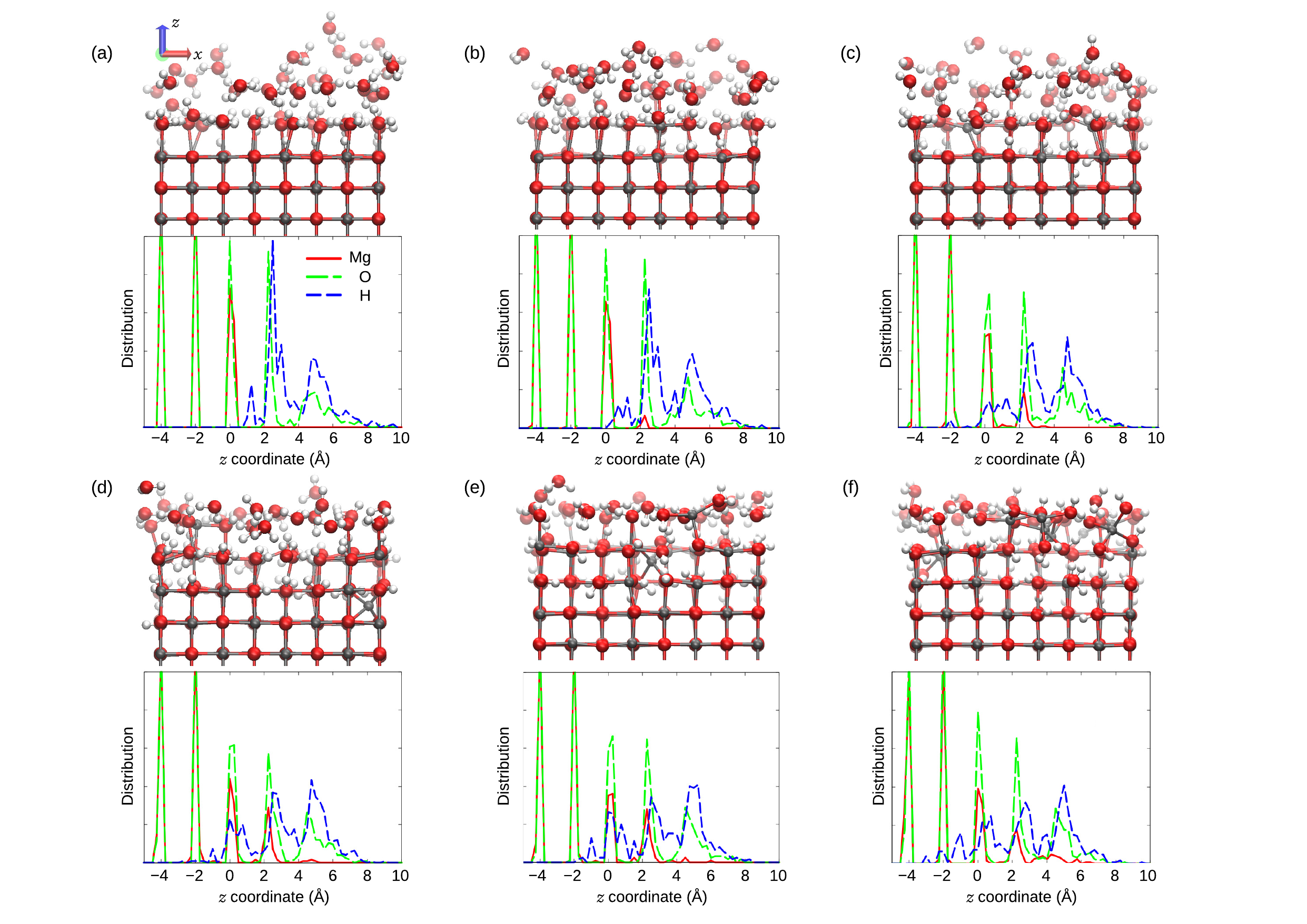}
    \caption{ \label{fig:overview_dens}
    Representative MgO/water interfacial structures optimized at the DFT level 
    (upper panels) and their corresponding $z$ coordinate (\AA) distributions 
    for Mg, O, and H atoms (lower panels) at various stages of the PS-MD cycle.
    The structures in the upper panels are optimized geometries from specific 
    PS-MD cycles: (a) 0, (b) 45, (c) 92, (d) 162, (e) 205, and (f) 310.
    The distributions in the lower panels are averaged over five structures 
    within the respective PS-MD cycle ranges: (a) 0–32, (b) 41–47, (c) 88–97, 
    (d) 156–162, (e) 201–205, and (f) 306–310.
    In the upper panels, Mg, O, and H atoms are colored gray, red, and white, respectively,
    while, in the lower distribution plots, the lines for Mg, O, and H are red, green, 
    and blue, respectively. 
    Both panels display the interface above the second MgO layer from the bottom. 
    Bonds between Mg and O atoms are shown for distances shorter than 2.3 \AA. 
    The $z$ coordinate is measured relative to the peak position of the outermost MgO layer.
    }
  \end{center}
\end{figure}

First, we overview the MgO/water interfacial structural changes 
obtained from the calculations. 
Figure \ref{fig:overview_dens} shows results from a representative run: 
the interfacial structures (upper panels) and the distributions of the $z$ coordinate 
for Mg, O, and H atoms (lower panels) with respect to the PS-MD cycle. 
The distributions calculated from the other four runs are shown in Figure \ref{figS:z-coord}. 
In the initial state (PS-MD cycle $\sim$ 0), the presence of two water monolayers is 
evident from the two distinct peaks in the oxygen distribution at $z > 0$ 
(Figure \ref{fig:overview_dens}a).
The first water layer on the MgO surface ($z$ $\sim$ 2.25 \AA) is shown by a very sharp 
peak, while the second water layer in contact with the gas phase is extended 
over a wide range in the $z$ direction. 
This densely packed first layer in the $z$ direction is consistent with previous 
studies\cite{Laporte2015,Ding2021}.
The orientation of the water molecules in the first layer can be roughly seen 
from the distribution of hydrogen. 
The hydrogen peak appears to be at almost the same position as the oxygen peak, 
indicating that many water molecules orient their molecular planes 
parallel to the MgO surface. 
The hydrogen peak at the slightly higher $z$ coordinate represents water molecules 
with an HB to molecules in the second water layer, and another hydrogen peak 
between the top surface of the solid MgO and the first water layer ($z$ $\sim$ 1 \AA) 
represents H atoms (or protons) dissociatively adsorbed on the MgO surface. 
Around the 50th PS-MD cycle (Figure \ref{fig:overview_dens}b), the small Mg peak appears 
at the first water layer, indicating that there are \mgion{} ions dissolved 
from the surface. 
Therefore, we found that the dissolution of \mgion{} ions is the first elementary 
process following the dissociative adsorption of water molecules. 
Around the 100th PS-MD cycle (Figure \ref{fig:overview_dens}c), as the dissolved 
\mgion{} ions increase, protons penetrate into the solid, which implies that the two 
molecular processes are correlated to each other. 
From the $\sim$150th PS-MD cycle (Figure \ref{fig:overview_dens}d), we can see 
the \mgion{} ions beginning to dissolve in the second water layer. 
The layer structure throughout the MgO/water interface is maintained even after 
the $\sim$150th PS-MD cycle (Figure \ref{fig:overview_dens}e and \ref{fig:overview_dens}f)
although there are minor changes in the distribution.
As these structural changes progress, the potential energy of the system generally 
decreases (see Section 2 in \SI), which is in qualitative agreement with the exothermic 
nature of MgO hydration. 
Further details in \SI{} show that the system eventually reaches an approximately 
64\% hydrated state from a potential energetic standpoint.
%
%

\begin{figure}
  \begin{center}
    \includegraphics[width=1.0\linewidth,trim=0cm 5cm 0cm 0cm, clip]{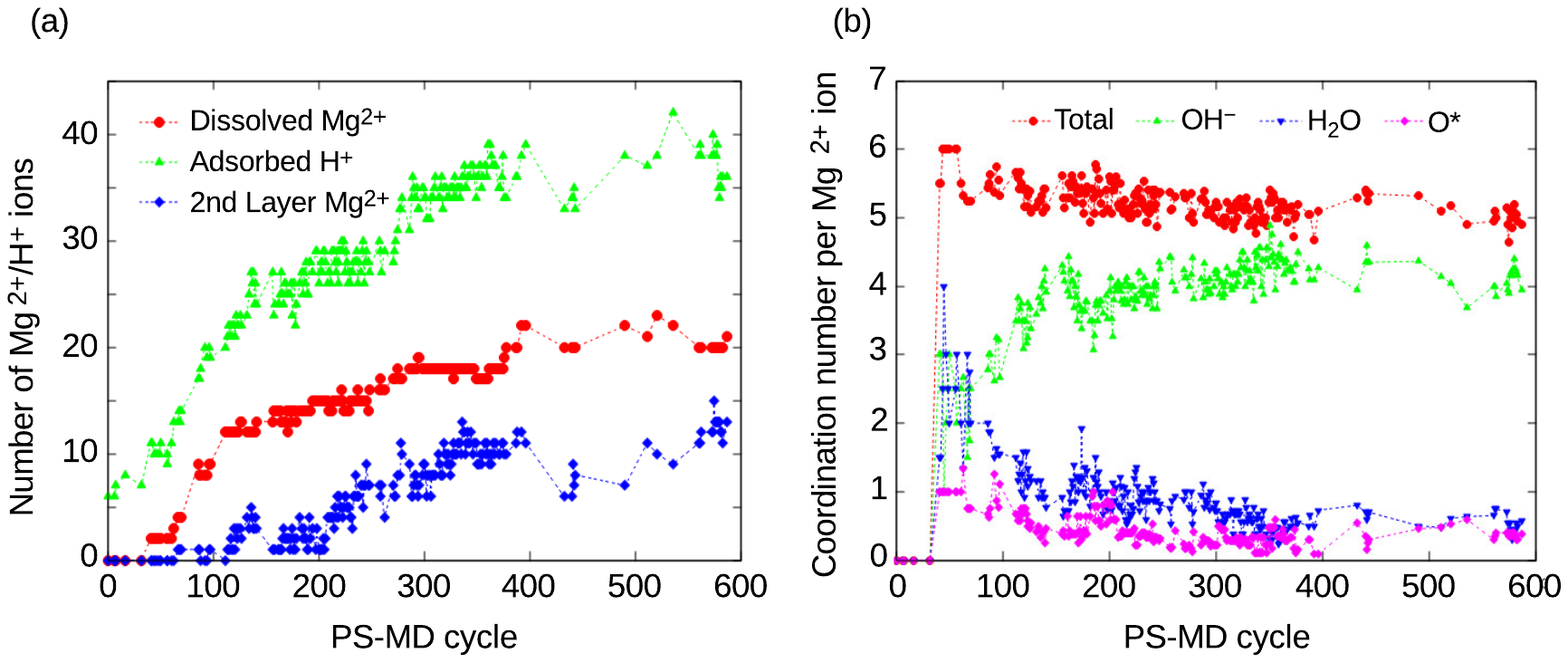}
    \caption{ \label{fig:overview_num}
    (a) Number of dissolved \mgion{} ions (red) and adsorbed/penetrated 
    protons (green) along the PS-MD cycle. 
    The number of \mgion{} ions in the second water layer (blue), 
    whose $z$ coordinates are greater than 2.75 \AA, is also shown.
    (b) Coordination number of dissolved \mgion{} ion along the PS-MD cycle. 
    The total coordination number is shown in red, and contributions to that 
    from OH$^-$ (green), H$_2$O (blue), and MgO surface oxygen atoms 
    (O$^*$; magenta) are also shown.
    The dotted lines are guides for the eye. 
    Similar results from the other four runs are presented 
    in Figures \ref{figS:dMg_pH_others} and \ref{figS:ngroups_others}. 
    Note that data points are not shown for PS-MD cycles where the optimized structure is
    less stable than the initial one (see Section 2 in \SI{} for details).
}
  \end{center}
\end{figure}

To quantitatively evaluate these interfacial rearrangements, 
we count the number of the respective ions along the PS-MD cycle. 
Figure \ref{fig:overview_num}a shows the changes in the number of \mgion{} ions dissolved 
from the MgO surface and the number of protons (H$^+$) adsorbed on (or penetrated into) 
the MgO solid. 
The dissolved \mgion{} and adsorbed protons are defined as ions at $z$ > 1.25 \AA{} and 
at $z$ < 1.75 \AA, respectively. 
The number of the protons at the PS-MD cycle zero indicates the number of water molecules 
dissociated in the initial state. 
We find from the figure that the dissolved \mgion{} ions increases monotonically along 
the PS-MD cycle and the protons are generated roughly in parallel with \mgion{} dissolution. 
The number of penetrated protons is approximately twice that of the \mgion{} ions. 
This is a result that the charges of the \mgion{} ions in the solid have been compensated 
by the protons, suggesting that the proton penetration correlates with the \mgion{} 
dissolution process. 
This observation appears to be consistent with a suggestion by Wogelius \etal{} 
who proposed the solid-state reaction mechanism in nucleation and growth\cite{Wogelius1995}. 
In the particularly early cycle (before the $\sim$100th PS-MD cycle), many \mgion{} 
ions are dissolved during the short period owing to the initially hydroxylated sites 
on the MgO surface. 
After the $\sim$100th PS-MD cycle, when about one thirds of the surface \mgion{} ions 
have been dissolved, the dissolved \mgion{} ions on the surface begin to migrate 
into the second water layer (blue plot in Figure \ref{fig:overview_num}a). 
This migration reduces the ions on the surface, which probably induces 
further dissolution of \mgion{} ions from the surface. 

%
%

To examine how the dissolved \mgion{} ions are stabilized, we then analyze the evolution 
of their local coordination environment. 
Figure \ref{fig:overview_num}b shows the change in oxygen species, OH$^-$ ion (green), 
H$_2$O molecule (blue), and surface oxygen atom O$^*$ (magenta), coordinated to the dissolved 
\mgion{} ions. 
We found that the dissolved \mgion{} ion is typically coordinated with more than 
five oxygen atoms (red plot in Figure \ref{fig:overview_num}b). 
At an early stage ($\sim$50th PS-MD cycle), the dissolved \mgion{} ions are stabilized by both 
OH$^-$ and H$_2$O species at equal ratio. 
Note that, at this stage, a surface oxygen always coordinates to the dissolved \mgion{} ions 
because they reside on the MgO surface. 
As the PS-MD cycle proceeds, the coordination of H$_2$O molecules is replaced with 
OH$^-$ ions. 
Eventually ($\sim$300th PS-MD cycle), the OH$^-$ ion bears $\sim$80\% of the total 
coordination. 
The increase in the coordination by OH$^-$ ions is definitely related to the consecutive 
water dissociation that occurs at the MgO/water interface (see Figure \ref{figS:O-dist} for 
the $z$ coordinate distributions of oxygen species). 
The observation that OH$^-$ ions form at the interface is also reported 
by On$\mrm{\check{c}\acute{a}}$k \etal{}\cite{Oncak2015}.

Based on the results in this section, the obtained structures can be classified qualitatively 
into three stages according to the Mg distribution change (Figure \ref{fig:overview_dens}): 
(1) the initial stage where water molecules are dissociatively adsorbed and no \mgion{} ions 
are dissolved (around the 0th PS-MD cycle), 
(2) the stage where \mgion{} ions are dissolved into the first water layer on the MgO surface 
(around the 50th PS-MD cycle), and 
(3) the stage where \mgion{} ions permanently exists in the second water layer  
(after around the 100th PS-MD cycle). 
In the following sections, we examine the interfacial structures of these three stages in detail, 
using all optimized structures obtained from the five independent runs.

\subsection{Dissociative adsorption of water molecules} \label{sec:results_water}
\begin{figure}
  \begin{center}
    \includegraphics[width=1.0\linewidth,trim=0cm 5cm 0cm 4cm, clip]{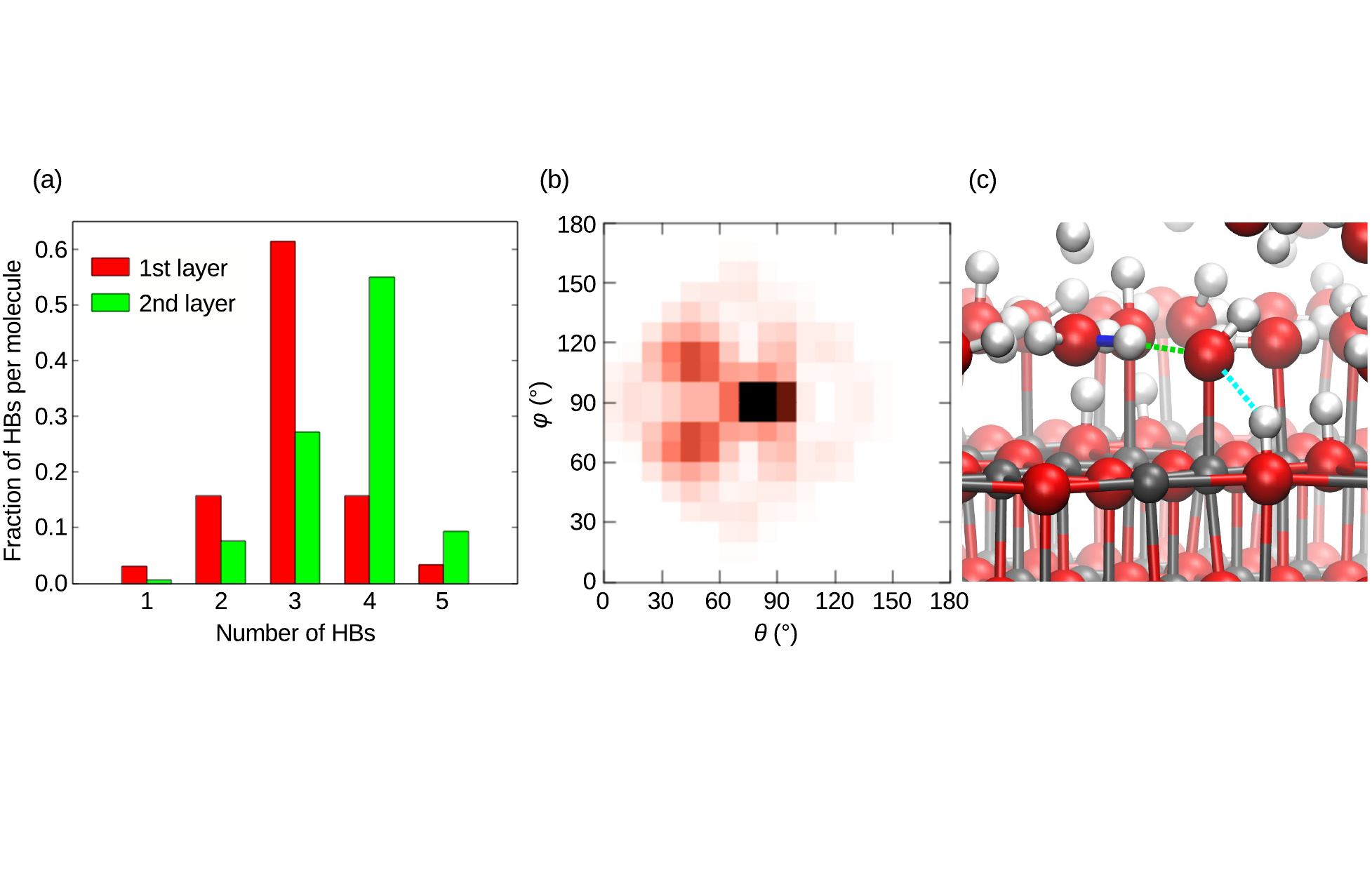}
    \caption{ \label{fig:water}
     (a) Fraction of the number of hydrogen bonds (HBs) per molecule in the first (red) and 
     second (green) water layers. 
     (b) Two-dimensional probability distribution of the orientation angles 
     $\theta$ ($^{\circ}$) and $\varphi$ ($^{\circ}$), for water molecules in the first layer,
     where a darker color indicates higher probability.
     (c) Key hydrogen-bonded configuration involved in water dissociation.
     In panel (c), the blue bond represents the elongated O-H bond (1.05 \AA), 
     while the green (1.54 \AA) and cyan (2.20 \AA) dotted lines represent the HBs 
     within the H$_2$O$\cdots$OH$^-$$\cdots$HO$^*$ complex. 
     Note that the oxygen and hydrogen atoms connected by the cyan dotted line originally 
     formed the downward O-H bond of a single water molecule before its dissociation. }
  \end{center}
\end{figure}

First, we investigate the structure and dissociative adsorption of water at the first stage, 
which represents the initial state of the MgO hydration reaction and serves as the precursor 
to \mgion{} dissolution. 
Our primary focus is to understand the dissociation of water molecules on the basis of 
the HB networks at the interface and the orientation of water molecules. 
The dissociation ratio of the surface water molecules was calculated to be 20\%. 
While various dissociation ratios have been reported in the literature, ranging from 0\% to 
75\%\cite{Odelius1999,Oncak2015,Oncak2016,Sassi2024,Carrasco2010}, 
our result is in good agreement with the result of \textit{ab initio} molecular dynamics 
by Ding and Selloni (22\%)\cite{Ding2021}. 
Figure \ref{fig:water}a shows the distribution of the number of HBs per water molecule. 
Hydrogen bonds were defined using the geometric criteria by Luzar and Chandler\cite{Luzar1996}. 
Water molecules in the first layer on the MgO surface (represented by red) predominantly 
form three HBs, fewer than in bulk water (3.7 HBs per molecule)\cite{Ashu2016,Kim2020}.
This difference can be attributed to the fact that these water molecules reside 
directly on the MgO surface. 
The oxygen atoms of the interfacial water molecules can be positioned atop Mg sites 
(Figure \ref{figS:1st_water_layer})   
because the distance between nearest-neighbor surface Mg (Mg$^*$) atoms ($\sim$2.98 \AA) 
is roughly similar to the typical oxygen-oxygen distance in hydrogen-bonded water molecules
($\sim$2.78 \AA). 
Also, the average distance between these Mg$^*$ atoms and the O atoms of the water molecules 
is $\sim$2.2 \AA, 
indicating a strong electrostatic interaction. 
Therefore, the coordination environment of the first-layer water molecules effectively 
consists of three HBs and one Mg$^*$-O bond. 
While this four-fold coordination agrees well with that of bulk water, 
the geometric structure is significantly different; 
Bulk water generally favors a tetrahedral coordination geometry, 
whereas the tetrahedral geometry for these first-layer molecules is highly distorted 
(Figure \ref{figS:tetrahed_OHorient}a).

This distorted structure of interfacial water is associated with their specific orientations 
on the interface. 
Figure \ref{fig:water}b shows a two-dimensional probability distribution based on two angles: 
$\theta$, the angle between the dipole moment vector of the water molecule and the $z$ axis 
(normal to the MgO surface), and $\varphi$, the angle between the vector connecting the two 
hydrogen atoms of the water molecule and the $z$ axis. 
The most prominent peak corresponds to ($\theta$, $\varphi$) $\sim$ (90$^{\circ}$, 90$^{\circ}$), 
indicating a dominant lying-down orientation where the plane of the water molecule is oriented 
parallel to the MgO surface. 
In addition, weaker peaks are observed around ($\theta$, $\varphi$) $\sim$ 
(50$^{\circ}$, 70$^{\circ}$) and ($\theta$, $\varphi$) $\sim$ (50$^{\circ}$, 110$^{\circ}$). 
These represent upward orientations, where the dipole moment is directed 
toward the gas phase ($\theta$ < 90$^{\circ}$). 
Water molecules with the lying-down orientation participate in a two-dimensional HB network 
within the first water layer, whereas those with the upward orientation participate 
in the two-dimensional HB network while also forming HBs with molecules in the second water layer.
Such HB networks and molecular orientations imply that specifically oriented 
(i.e., downward-oriented) water molecules dissociate.
Interestingly, the downward orientation, where the dipole moment points toward the MgO 
surface, was rarely observed (Figure \ref{fig:water}b). 
This scarcity is likely due to the dissociation of these downward-oriented molecules, 
triggered by strong interactions with their neighbors. 
Indeed, the resulting hydroxide ions, similar to the water molecules, adopt 
a lying-down or upward orientation (Figure \ref{figS:tetrahed_OHorient}b).
That is, it is water molecules adopting a downward orientation, for instance 
due to thermal fluctuations, that are promoted to dissociate by the surrounding 
HB network.
This dissociation mechanism is consistent with the observed interatomic distances and 
configurations. 
The average O-H bond length of the interfacial water molecules was longer (1.003 \AA) 
than that of the isolated water molecule (0.979 \AA)  (Figure \ref{figS:OH_bond_and_rdf}a), 
suggesting that they are primed for dissociation. 
Critically, we observed short and strong HBs between dissociated hydroxide ions and 
adjacent water molecules (Figure \ref{figS:OH_bond_and_rdf}b). 
These HBs appear in configurations, typified by Figure \ref{fig:water}c, 
where an adsorbed proton is located beyond the HB between the hydroxide ion and 
the water molecule (green dotted line).
Such configurations strongly suggest that the downward O-H bond dissociates 
(the proton is pushed out) by the strong HB donor of a neighboring water molecule 
from the opposite side. 

On the other hand, the second water layer on the gas phase seems to 
possess a water structure more similar to bulk water than the first layer. 
Figure \ref{fig:water}a shows that, in the second layer, the average number of HBs 
per water molecule is 3.7, which is quite similar to that of bulk water (3.7). 
The orientation of water molecules appears largely random although the downward
orientation is slightly increased due to the HB with molecules in the first layer 
(Figure \ref{figS:orientation_2nd_layer}). 
The tetrahedrality is also considerably closer to that 
of bulk water compared to the first layer (Figure \ref{figS:tetrahed_OHorient}a). 
In fact, these characteristics have been suggested in previous studies
\cite{McCarthy1996,Laporte2015}. 
Nevertheless, a number of molecules are found to deviate from the tetrahedral 
structure given that there are water molecules with small or even negative 
tetrahedral order parameters (Figure \ref{figS:tetrahed_OHorient}a). 
The deviation from the tetrahedral structure likely results from the molecules' effort 
to maintain HBs even on the gas phase. 
While Laporte \etal{} described the water in the second layer as the bulk-like water 
because the O-O distance in hydrogen bonded pairs is quite close to that of bulk water
\cite{Laporte2015}, our results suggest that distorted structures allow the molecules 
to maintain such O-O distances even in the different environment from the bulk. 
Note that, in contrast, Ding \etal{} suggested on the basis of the molecular diffusion 
that the bulk-like behavior begins from the third layer\cite{Ding2021}.
%
%

\subsection{Dissolution of \mgion{} ions} \label{sec:results_dissolv}
\begin{figure}
  \begin{center}
    \includegraphics[width=1.0\linewidth,trim=0cm 4cm 0cm 3cm, clip]{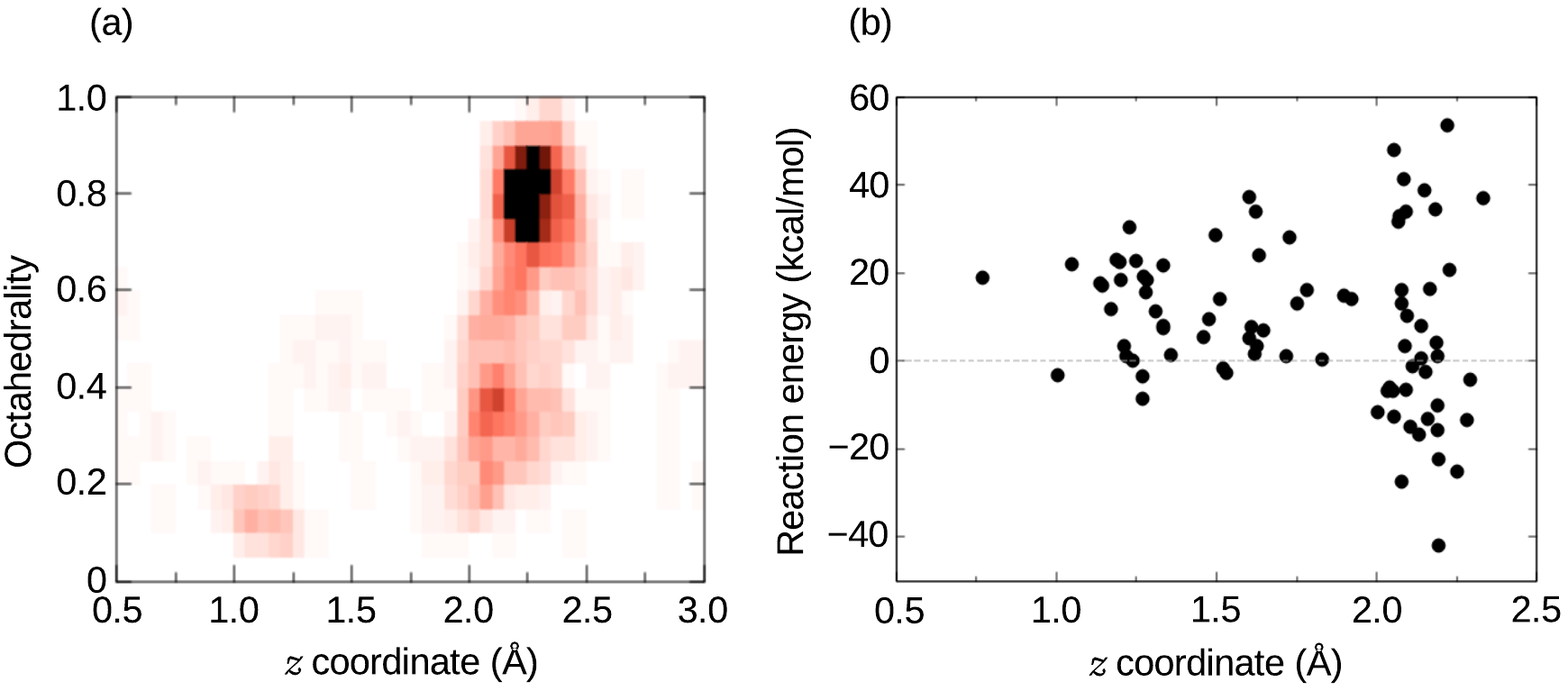}
    \caption{ \label{fig:octa_reaceng}
     (a) Two-dimensional probability distribution of the octahedral order parameter
     \cite{Zimmerman2015} for the solvated \mgion{} complex as a function of its 
     $z$ coordinate calculated by the five independent runs. 
     A darker color indicates higher probability. 
     (b) Reaction energy (kcal/mol) plotted against the $z$ coordinate of the \mgion{} 
     ion obtained from the nudged elastic band calculations.
     Only dissolution processes with a locally stable product structure are used.
}
  \end{center}
\end{figure}

The second stage spans from the first observation of the \mgion{} dissolution process
until just before \mgion{} ions are constantly present in the second water layer. 
In this period, roughly 10 Mg$^*$ atoms (about 30\% of the surface Mg atoms) 
are dissolved (Table \ref{tabS:stage2}). 
In this section, we mainly examine the initial \mgion{} dissolution process. 
According to the structures optimized at the DFT level, 
the dissolved \mgion{} ions are typically located on top of O$^*$ sites. 
As depicted in Figures \ref{fig:overview_dens}b and \ref{fig:overview_dens}c, 
their $z$ coordinates ($\sim$2.25 \AA) are comparable to those of the oxygen atoms 
in the first water layer. 
This observation that most of the dissolved \mgion{} ions are found above O$^*$ sites 
implies that the fully hydroxylated MgO surface during the first stage hinders 
the subsequent elementary process of \mgion{} dissolution. 
Most of these \mgion{} ions on top of O$^*$ sites exhibit an octahedral six-coordinate 
structure (Figure \ref{fig:octa_reaceng}a).
These solvated ions include what On$\mrm{\check{c}\acute{a}}$k \etal{} 
described as a contact ion pair\cite{Oncak2015}.
On the other hand, a few dissolved \mgion{} ions are observed bridging two O$^*$ sites at 
$z$ $\sim$1 \AA{} with the low octahedrality, 
making them unstable compared to the five- or six-coordinate \mgion{} ions. 
(see below for details). 
Note that in the present calculations no H$^+$ penetration was found before \mgion{} 
dissolution in contrast to the suggestions by previous works\cite{Wogelius1995,Oviedo1998}.
%
%

Our focus in the second stage is what causes the exothermic dissolution of \mgion{} ions. 
Unfortunately, however, the structures optimized at the DFT level represent only fragmented 
local stable points, and it is difficult to examine the reaction pathway and its corresponding 
energy profile. 
Therefore, we performed nudged elastic band (NEB) calculations to determine the dissolution 
pathway for each surface \mgion{} ion,
using the final configuration from the first stage as the reactant.
The calculation generates a total of 160 independent dissolution pathways, arising from
32 surface \mgion{} ions for each of the five independent runs 
(see \SI{} for the computational details). 
We counted the \mgion{} ion with the $z$ coordinate greater than 0.75 \AA{} in the product 
as the successfully dissolved \mgion{} ion. 
This criterion was met by 83 \mgion{} ions, while for the remaining 77 ions, 
no stable dissolved state was found; the latter ions returned to the solid surface. 
This observation indicates that the successful dissolution process is highly dependent on 
the local environment of water molecules. 
Furthermore, among the 83 samples where the dissolution was observed, only 25 samples exhibited 
an exothermic (negative) reaction energy. 
The resulting energy profiles of the dissolution processes are presented 
in Figure \ref{figS:neb_profile}. 
For the dissolution processes with a negative reaction energy, the average reaction barrier 
relative to the reactant was found to be $\sim$12$\pm$12 kcal/mol. 
This barrier height is comparable to the activation energy of $\sim$13-18 kcal/mol reported 
in previous studies\cite{Smithson1969,Thomas2014,Kuleci2016}  
that suggested \mgion{} dissolution or interfacial reactions as the rate-determining process. 
The significant variation of the calculated barrier height reflects the severely 
heterogeneous nature of this process. 
Also, this barrier is not so high, indicating that the dissolution can occur exothermically 
even in the perfect (defect-free) MgO (100) surface, not necessarily at the edge.

The above observation raises questions of which product structures 
favor exothermic dissolution and what drives this process. 
Our analysis revealed a correlation between the reaction energy and the $z$ coordinate 
of the dissolved \mgion{} ion (Figure \ref{fig:octa_reaceng}b). 
Specifically, the dissolution process tends to be exothermic when the \mgion{} ion is 
positioned 2.0 \AA{} or greater from the surface. 
Given that \mgion{} ions in this region tend to adopt an octahedral structure 
(Figure \ref{fig:octa_reaceng}a), the product is expected to be stabilized 
by sufficient coordination. 
Thus, for the dissolution to be energetically favorable, it is crucial that the \mgion{} ion 
positions itself directly above an O$^*$ site to form the octahedral structure. 
However, such \mgion{} ion's position and an octahedral structure 
do not guarantee a negative reaction energy. 
Therefore, we further classified the product structures where the $z$ coordinate of 
the \mgion{} ion is 2.0 \AA{} or greater.

\begin{figure}
  \begin{center}
    \includegraphics[width=1.0\linewidth,trim=1cm 0cm 2cm 0cm, clip]{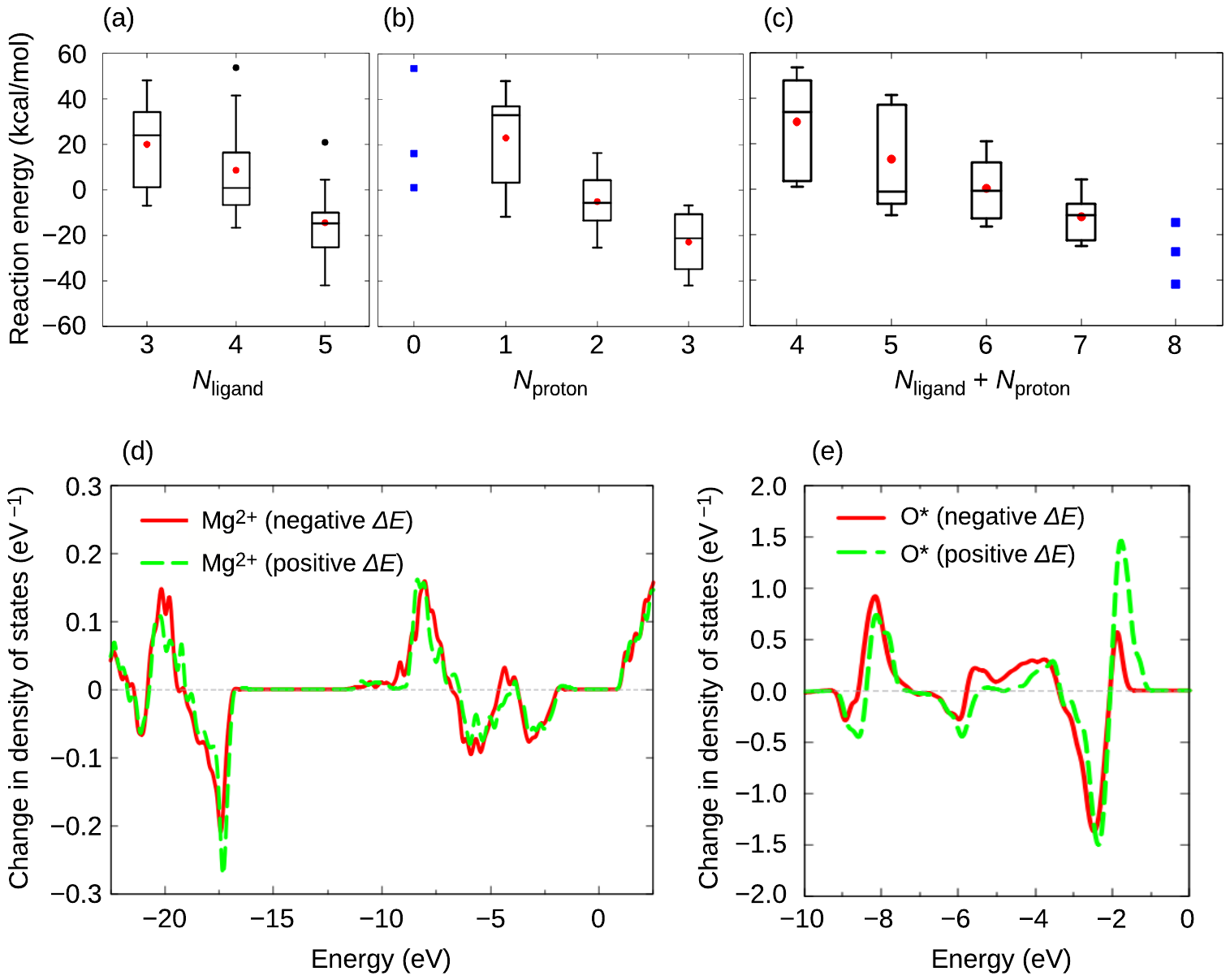}
    \caption{ \label{fig:dissolution}
    (Upper) Boxplot of the reaction energy ($\Delta E$, kcal/mol) for the \mgion{} 
    dissolution process classified by (a) the number of ligands (OH$^{-}$ ions and 
    H$_2$O molecules) coordinated to the dissolved \mgion{} ion, $N_{\mrm{ligand}}$, 
    (b) the number of protons adsorbed on O$^*$ atoms adjacent to the vacancy generated 
    by the dissolution, $N_{\mrm{proton}}$, 
    and (c) the sum of the number of the ligands and protons. 
    The distance cutoffs for \mgion{}-O coordination and O$^*$-H$^+$ bonding are set to 
    2.5 \AA{} and 1.2 \AA, respectively.
    The red and black circles indicate mean reaction energy and outlier in the category, 
    respectively, and the blue squares are used when there are less than four samples.
    (Lower) Change in the projected density of states of the dissolved \mgion{} ion (d) and 
    the O$^*$ atoms originally coordinated to the dissolved \mgion{} ion (e) 
    for the dissolution process. 
    The density of states of the Mg and O$^*$ atoms are projected onto their (3s, 3p) and 
    (2s, 2p) orbitals, respectively. 
    The changes according to the process with negative and positive reaction energies are 
    colored red and green, respectively. 
    In both panels, the reference energy is set to the vacuum level in the slab model.
}
  \end{center}
\end{figure}

Figure \ref{fig:dissolution}a presents a box plot of the reaction energy classified 
by the number of ligands (H$_2$O or OH$^-$) coordinated to the dissolved \mgion{} ion. 
It should be noted that since the dissolved \mgion{} ion is positioned on an O$^*$ atom, 
the coordination number of five in the figure denotes a total coordination number of six, 
including the O$^*$ atom. 
Although the variation in the reaction energy is quite substantial, both the mean 
(red circle) and median (horizontal bar in the box) values of the reaction energy 
tend to decrease as the number of ligands increases. 
This indicates that the dissolved \mgion{} ion is stabilized by coordination with 
water molecules and hydroxide ions. 
However, even with total six-coordination allowing for an octahedral structure, 
a negative reaction energy is not guaranteed. 
Conversely, configurations with a total coordination number of four or five can still yield 
negative reaction energies. 
This raises the next question; is there another structural feature of the products 
that strongly correlates with the reaction energy? 
The stability of the product also depends on how significantly a vacancy, 
which is created at the original position of the dissolved \mgion{} ion, is stabilized. 
The role stabilizing the vacancy is played by protons adsorbed on nearby O$^*$ sites. 
Figure \ref{fig:dissolution}b classifies the reaction energy according to the number of 
these protons surrounding the vacancy. 
Again, we can observe a trend that the reaction energy decreases as the number of the 
protons increases. 
According to formal charge considerations, two protons can largely neutralize 
the electrostatic instability (i.e., repulsive interaction between the O$^*$ atoms 
around the vacancy) caused by \mgion{} dissolution. 
On$\mrm{\check{c}\acute{a}}$k and co-workers called this neutralizing process 
"defect healing"\cite{Oncak2016}.
Indeed, a rough boundary between the positive and negative reaction energies seems to lie 
between having one and two protons. 
Notably, within our samples, all product structures with three protons infilling the vacancy 
show negative reaction energies. 
Compared to the classification by the coordination number of the \mgion{} ion 
(Figure \ref{fig:dissolution}a), the classification by the number of surrounding protons 
exhibits smaller variance and fewer outliers relative to the median. 
Therefore, we suggest that the stabilization of the vacancy is particularly crucial 
for achieving a favorable dissolution structure. 
In addition, once an \mgion{} ion dissolves and the vacancy is occupied by protons,
the ion's return to the surface becomes infeasible; the process is effectively irreversible.
Finally, Figure \ref{fig:dissolution}c classifies the reaction energy according to 
the combined count of the coordinating ligands and the protons infilling the vacancy. 
As expected, the clearer trend can be found that the stability of the product increases 
with a larger total number of these species.
%
%

DFT calculations not only provide plausible molecular interfacial structures 
but also offer insights into the electronic states. 
Figure \ref{fig:dissolution}d shows how the projected density of states (pDOS) of 
the dissolving \mgion{} ions ($z$ > 2.0 \AA) changes in the dissolution process. 
(The pDOS changes of the O atoms coordinating to a dissolving \mgion{} ion are
shown in Figure \ref{figS:water_O_diffDOS}.)
The red solid and green dashed lines correspond to the changes in pDOSs obtained 
from the dissolution processes with negative and positive reaction energies, respectively. 
We found that high-energy occupied states at about $-5.5$ eV and $-3.0$ eV diminish 
and largely shift to a lower energy region around $-8.0$ eV. 
This is a consequence of the change in the chemical state of the \mgion{} ion 
as its coordination changes from bonding with O$^*$ atoms to coordination with H$_2$O 
and OH$^-$ ligands (Figure \ref{figS:reac_prod_O_dos}). 
This stabilization of the one-electron state can be one of the driving forces for 
the dissolution process (cf. Figure \ref{fig:dissolution}a). 
However, this feature is almost unchanged between exothermic and endothermic 
dissolution processes. 
Figure \ref{fig:dissolution}e shows, on the other hand, the change in pDOSs 
of the O$^*$ atoms originally coordinated to the dissolving \mgion{} ion. 
Although similar pDOS changes appear to be observed regardless of the reaction energy, 
a significant difference is apparent around $-1.9$ eV. 
In the case of processes with a positive reaction energy, the pDOS at $-$1.9 eV
substantially increases (green dashed line), whereas the increase is considerably 
smaller for a negative reaction energy (red solid line). 
In the former case, a large portion of the state at $-2.5$ eV is found to shift to 
$-1.9$ eV, which corresponds to the dissolution process where the O$^*$ atoms fail 
to be stabilized by the proton adsorption. 
The stabilized O$^*$ states, in contrast, can be seen around a lower energy region 
($-6$ eV $\sim$ $-4$ eV), which is related to the dissolution process with a negative 
reaction energy. 
Therefore, this increase in the pDOS at $-1.9$ eV reflects the fact that 
the O$^*$ atoms are not stabilized by protons during the dissolution process. 
These results indicate that the suggestion from Figure \ref{fig:dissolution}a-c is 
also supported from the viewpoint of electronic states; 
the proton adsorption on the O$^*$ atoms plays an important role 
in the exothermic dissolution process.

Understanding the exothermicity from a viewpoint of the reactant structure is not 
straightforward in the \mgion{} dissolution process. 
Since the product structure infilling the vacancy is expected to be intimately related 
to the hydroxylation around the Mg$^*$ atom to be dissolved in the reactant state, 
the latter would be linked to the exothermicity of the reaction. 
However, we found no significant correlation between the number of hydroxylated O$^*$ atoms 
and the reaction energy (Figure \ref{figS:boxplot_reactant}). 
This is because the slight displacement of the Mg$^*$ atom significantly changes 
the orientation of surrounding hydroxide ions and water molecules, and consequently 
triggers water dissociation or proton transfer. 
Indeed, Hu and co-workers have reported that the energy barrier is small for proton transfer 
among water molecules and hydroxide ions on the MgO surface\cite{Hu2010}.
Therefore, our results indicate that \mgion{} dissolution is a concerted process 
coupled with rearrangements of the local HB network.
This implies that a detailed analysis of the HB network is necessary to understand 
the exothermic \mgion{} dissolution from a viewpoint of the reactant structure.
According to the present static calculations, the range of proton transfer is limited 
to the second hydration shell of the \mgion{} ion, suggesting that an analysis of 
a relatively local network is sufficient. 
This static picture is potentially modified by dynamic fluctuations, 
which could extend the influence from the \mgion{} ion displacement.

\subsection{Nucleation and growth of \mgoh} \label{sec:Results_nucleation}

The third stage of the present hydration process represents states where dissolved 
\mgion{} ions are persistently located in the second water layer 
(i.e., after around the 100th PS-MD cycle). 
Here, we focus on the formation of Mg-OH chains relevant to the \mgoh{} nuclei 
and show the possibility of two previously suggested nucleation and growth mechanisms. 

One is the solid-state reaction mechanism, 
wherein an \mgoh{}(0001) layer forms on the MgO (111) surface. 
To investigate this mechanism, we examined proton penetration through vacancies 
created by \mgion{} dissolution. 
At approximately the 300th PS-MD cycle (the approximate limit for the structural 
analysis herein; see Section 2 in \SI{} for details), 
eight protons on average have penetrated into the first sublayer 
(Figure \ref{figS:penetrated_H}).
This quantity corresponds to one-quarter of the O atoms in that layer. 
These penetrated protons bind to O atoms, compensating for the charge of the dissolved 
\mgion{} ions. 
Consequently, the O atoms in the solid largely remain in their original sites, 
which is consistent with the early experimental study by Wogelius \etal{}\cite{Wogelius1995}. 
The resulting OH$^-$ ions are distributed on the surface and in the first sublayer 
(Figure \ref{figS:OH_row}), forming rows on the (111) plane of the MgO crystal that 
align along the [110] direction.
This configuration is consistent with the molecular hydration process 
proposed by Jug \etal{}\cite{Jug2007_path}. 
However, this process stalled almost completely at the first sublayer 
(Figure \ref{figS:penetrated_H}).
For proton penetration to proceed deeper, \mgion{} ions must dissolve from the second 
sublayer, but such dissolution was hardly observed.
This is because the \mgion{} ions in the second sublayer are strongly stabilized  
by the persistent MgO lattice framework (i.e., lattice potential dominated by 
Madelung potential).
Thus, it is highly unlikely that a large \mgoh{} film forms in the solid by the proton 
penetration as long as the MgO solid framework is maintained.
This implies that if a solid-state reaction mechanism were to be assumed, 
the extraction of internal \mgion{} ions would become the rate-determining process. 
This implication is consistent with the findings of Ishida and Ishimura\cite{Ishida2024}.

An alternative pathway using the MgO (111) plane may be to form [110] steps and 
gradually expose underlying (100)-oriented terraces.
This surface reconstruction requires the continuous dissolution of surface Mg atoms, 
which in turn allows for the construction of OH$^-$ rows in the newly exposed deeper regions
(a schematic illustration of this process is shown in Figure \ref{figS:sketch}).
However, the resulting structure would slightly differ from the stable \mgoh{} structure. 
For example, since the OH$^-$ rows are constrained by the strong lattice potential from
the underlying MgO(111) plane, they are forced to adopt an O-O distance of $\sim$2.98 \AA{} 
similar to that in the MgO crystal.
This distance is inconsistent with the $\sim$3.15 \AA{} distance found in a relaxed \mgoh{} 
structure. 
As long as the substrate maintains the MgO rock-salt structure, this lattice mismatch 
makes structural relaxation difficult, leading to energetically unfavorable and strained 
structures.

\begin{figure}
  \begin{center}
    \includegraphics[width=1.0\linewidth,trim=0cm 6cm 0cm 3cm, clip]{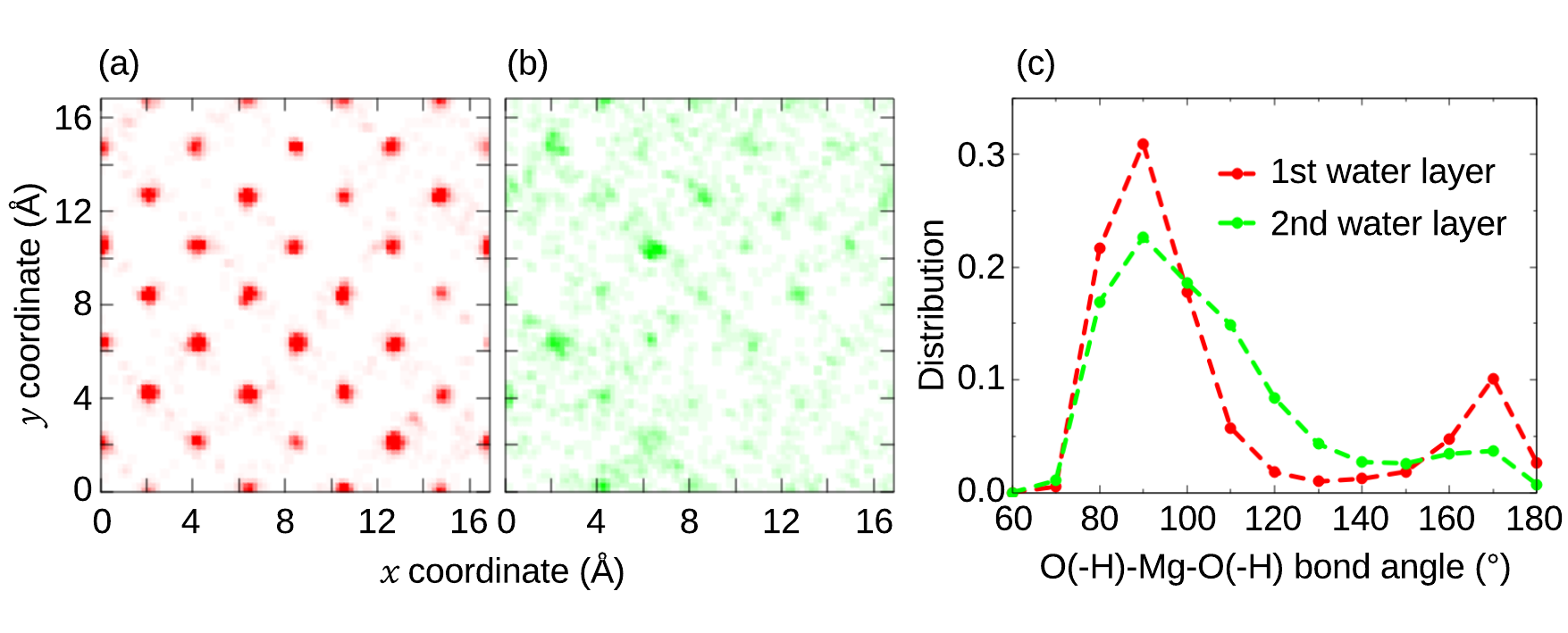}
    \caption{ \label{fig:map_and_angle}
    Normalized lateral distribution of dissolved \mgion{} ions in the (a) first 
    (1.0 \AA{} < $z$ < 3.0 \AA) and (b) second ($z$ > 3.0 \AA) water layers.
    (c) Bond angle distribution of O(-H)-Mg-O(-H) chains found in the first (red) 
    and second (green) water layers.  
    In both panels, the \mgion{} ions are sampled from snapshots in the third stage.
}
  \end{center}
\end{figure}

The other possible nucleation and growth is the formation of \mgoh{} via dissolution,
aggregation, and precipitation. 
When \mgion{} ions dissolve and move away from the MgO surface, 
they experience a weaker lattice potential. 
(On$\mrm{\check{c}\acute{a}}$k \etal{} seem to refer to these ions 
as a solvent-separated ion pair\cite{Oncak2015}.)
Figure \ref{fig:map_and_angle}a illustrates the lateral distribution of \mgion{} ions 
situated on the MgO surface, with the $z$ coordinates ranging from 1.0 \AA{} to 3.0 \AA. 
These ions are clearly well-aligned on O$^*$ sites with a strong influence 
from the underlying MgO lattice. 
These aligned ions can form structures corresponding to energetically stable stripe 
configurations described by On$\mrm{\check{c}\acute{a}}$k \etal\cite{Oncak2015,Oncak2016}
In contrast, Figure \ref{fig:map_and_angle}b shows that \mgion{} ions with the $z$ 
coordinates of 3.0 \AA{} or greater are distributed more randomly in the $xy$ plane. 
This is attributed to the fact that these \mgion{} ions are coordinated 
purely by water molecules or hydroxide ions. 
This difference in the local environment is also reflected in the local structure. 
Figure \ref{fig:map_and_angle}c shows the bond angle distribution for O(-H)-Mg-O(-H) chains 
observed in the third stage. 
The distribution for the chains on the MgO surface 
(red) exhibits sharp peaks around 90$^{\circ}$ and 170$^{\circ}$. 
This serves as evidence that the influence of the MgO rock-salt structure remains significant. 
We found from the green plot, obtained from the chains in the second water layer ($z$ > 3.0 \AA), 
that this strong influence from the MgO crystal diminishes. 
Although a peak is present at 90$^{\circ}$, its intensity is reduced, and structures
characterized by wider bond angles (100$^{\circ}$-120$^{\circ}$) become more prevalent. 
In addition, the peak at 170$^{\circ}$ changes to a shoulder. 
Such lateral positions and local structures surrounding \mgion{} ions 
that are less affected by the MgO lattice potential are more pronounced at the outermost 
interface (Figures \ref{figS:Mg_site_all} and \ref{figS:O-Mg-O_all}).

%
%

\begin{figure}
  \begin{center}
    \includegraphics[width=1.0\linewidth,trim=0cm 3cm 0cm 1cm, clip]{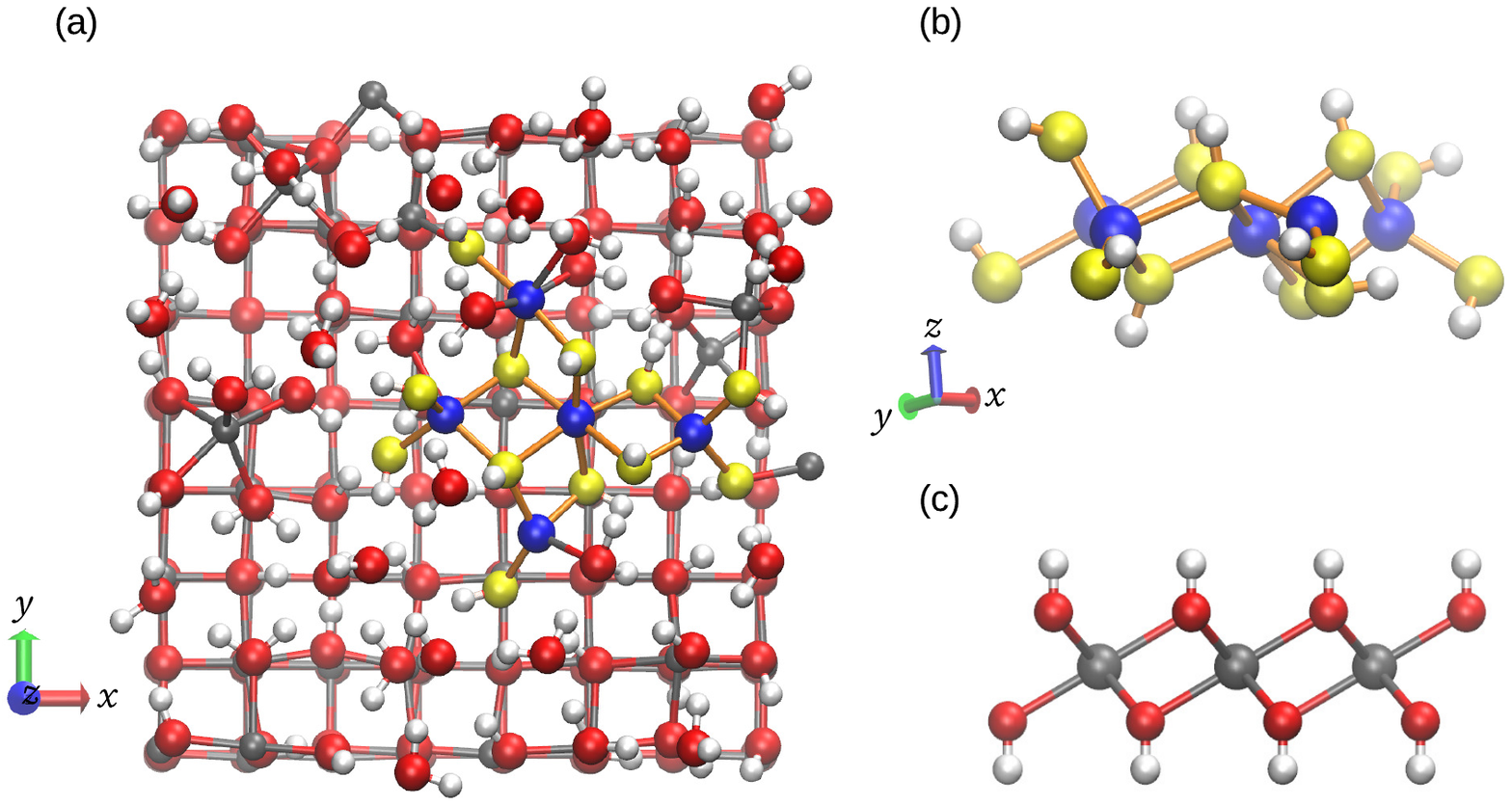}
    \caption{ \label{fig:mgoh_chain}
    (a) Representative MgO/water interfacial structure with an Mg-OH chain 
    far from the MgO surface. 
    The Mg and O atoms of the chain are highlighted by blue and yellow, respectively, 
    and the other Mg and O atoms are colored gray and red, respectively. 
    The H atoms are colored white. 
    (b) The Mg-OH chain seen from another angle, which is similar in morphology to part of 
    the \mgoh{} crystal displayed in panel (c).
}
  \end{center}
\end{figure}

Figure \ref{fig:mgoh_chain}a presents a typical example of Mg-OH chains found 
in the second water layer. 
The chain comprises five dissolved \mgion{} ions and they are interconnected solely 
by OH$^-$ ions, without any water molecules or surface oxygen atoms. 
In Figure \ref{fig:mgoh_chain}b, the Mg-OH chain is extracted from the interface and viewed 
from a different angle. 
This structure is found to have a morphology reminiscent 
of \mgoh{} crystals (shown in Figure \ref{fig:mgoh_chain}c).
This calculation result suggests, based on their structural similarity, that 
the Mg-OH chains in regions far from the MgO surface
can be precursors of \mgoh{} crystal nuclei.
Such Mg-OH chains are likely the surface clusters or amorphous phases experimentally 
observed on MgO crystals\cite{Lee2003,Rimsza2019,Luong2023_nanoscale,Bracco2024}. 

However, the Mg-OH chains found in the present calculation contain four- and five-coordinate 
\mgion{} ions, 
unlike the uniformly six-coordinate \mgion{} ions within the crystalline \mgoh{}. 
The primary reason for this incomplete coordination is the thinness of the water layer. 
For \mgion{} ions to dissolve from the MgO surface, achieve adequate solvation, and establish 
their preferred coordination by OH$^-$ ions, the space provided by an initial two-layer 
water film is insufficient \cite{Luong2023_Langmuir}. 
Indeed, the upper part of the Mg-OH chain shown in Figure \ref{fig:mgoh_chain}a is exposed 
to the gas phase, while some of its lower OH$^-$ ions are located on the MgO surface. 
Therefore, to gain further evidence for the mechanism of \mgoh{} crystal nuclei formation 
far from the MgO surface, 
we require additional simulations involving a more substantial number of water molecules.
%
%

\begin{figure}
  \begin{center}
    \includegraphics[width=1.0\linewidth,trim=0cm 8cm 0cm 3cm, clip]{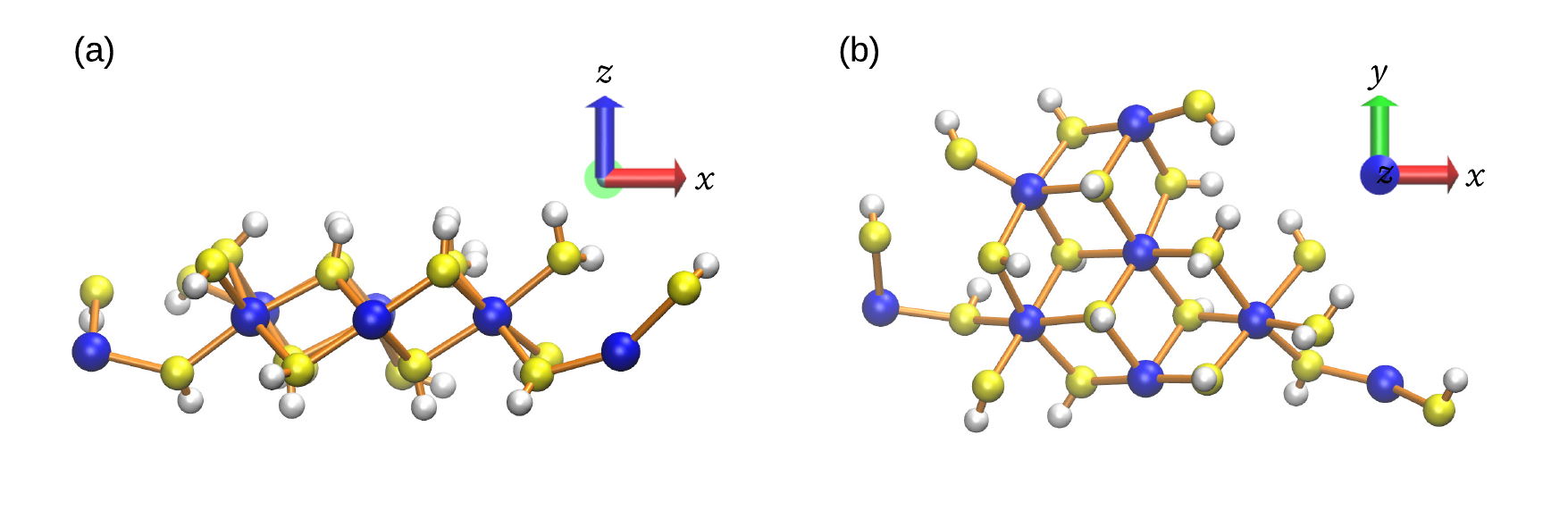}
    \caption{ \label{fig:mgoh_in_water}
    Side (a) and top (b) views of a representative Mg-OH chain obtained in water. 
    The Mg, O, and H atoms are colored blue, yellow, and white, respectively. 
    Surrounding molecules are omitted for clarity.
}
  \end{center}
\end{figure}

In order to demonstrate that the nucleation and growth of \mgoh{} occurs in an environment 
with sufficient water molecules, we performed the additional simulations mentioned above. 
We extracted an amorphous phase consisting of Mg, O, and H atoms on the MgO surface 
($z$ > 1 \AA), placed it in bulk water, and performed PS-MD simulations and subsequent 
geometry optimizations at the DFTB and DFT levels (see \SI{} for computational details). 
As a result, we successfully obtained crystalline Mg-OH chains exhibiting an \mgoh-like structure 
(Figure \ref{fig:mgoh_in_water}). 
We find from the figure that eight \mgion{} ions form a two-dimensional layered structure, 
and the bridging OH$^-$ ions are arranged in positions that precisely match those 
in the \mgoh{} crystal structure (Figure \ref{fig:mgoh_chain}c). 
This structure is considerably more well-ordered compared to the Mg-OH chains observed 
in the MgO/water interface system (Figure \ref{fig:mgoh_chain}b). 
The aqueous environment without the MgO lattice potential 
likely contribute to the formation of this well-ordered crystal nucleus. 
In the present calculations, the further growth of this \mgoh{} crystal nucleus was not 
observed. 
To achieve such growth, we would need more refined adjustments to the potential scaling 
parameter and longer simulation durations in the PS-MD simulations. 
Elucidating the nucleus growth through such simulations remains a task for future work.
Nevertheless, the present calculation results clearly support the mechanism that \mgoh{} 
crystal nuclei dominantly form via dissolution, aggregation, and precipitation 
in regions far from the MgO surface.
%
%

\clearpage
\subsection{Implications on the rate-determining process} \label{sec:results_rds}
Identifying the rate-determining process is a key to understanding 
the mechanism of complex and multi-step solid-surface chemical reactions.
Our investigation focused on two important processes in MgO hydration: 
the initial dissolution of an \mgion{} ion and the nucleation of \mgoh. 
Our calculations estimated the energy barrier for the dissolution of an 
\mgion{} ion from the defect-free MgO(100) surface to be approximately 12 kcal/mol. 
According to transition state theory, this barrier height corresponds to a timescale
of milliseconds. 
We did not estimate the barrier height for the subsequent \mgion{} dissolution process. 
However, in the presence of sufficient water, we expect the barrier height to be 
comparable to or lower than that for the initial one.
This is because a surface that has undergone \mgion{} dissolution is already 
disordered and defect-rich.
On the other hand, we expect that the nucleation of \mgoh{} from a supersaturated 
aqueous layer is a faster process than the \mgion{} dissolution. 
Because the PS-MD simulations for the nucleation in water were performed under a milder
potential scaling condition than that used for the structure generation of the MgO/water
interface (see Computational Methods in \SI{}), the energetic barrier for the nucleation 
is expected to be lower than that for the \mgion{} dissolution.
Considering the high concentration of dissolved \mgion{} ions in a water layer,
the entropic barrier in nucleation is expected to be minimal.
Therefore, the rate-determining process, i.e., the elementary process with 
the highest activation barrier in the overall reaction, is most likely 
the dissolution of \mgion{} ions.
This finding is consistent with the experimental results of Kuleci \etal, which suggest 
that the hydration rate is governed by interfacial reactions\cite{Kuleci2016}.
This implication leads to a seeming discrepancy between the millisecond timescale 
of this rate-determining process and the hour-long timescale of the macroscopic reaction
\cite{Kato1996,Thomas2014,Xing2018,Maltseva2019,Kondo2021,Luong2023_Langmuir}, 
but it is reconciled by the sequential nature of the reaction. 
The transformation of the MgO surface is not a single event but requires a long 
sequence of these dissolution processes to supply sufficient \mgion{} ions 
for the nucleation and growth of a new \mgoh{} phase. 
Therefore, while the individual rate-determining process is fast, the overall reaction 
rate is determined by the cumulative effect of these sequential dissolutions.
This consideration can explain the experimental observation that hydration is 
much faster on surfaces with a high density of steps and edges\cite{Thomas2014,Baumann2015}. 
These defect-rich sites simply lower the activation barrier of the rate-determining 
process itself. 
This increases the rate of the individual events, which in turn accelerates 
the overall sequence and, consequently, the macroscopic reaction rate.

Note that some experimental studies suggest the diffusion of \mgion{} ions 
and water molecules through the product \mgoh{} layer on the MgO surface can become 
rate-limiting\cite{Lee2003,Tang2018,Xing2018,Luong2023_nanoscale,Pettauer2024}. 
This process, however, governs the later stages of the hydration after the initial 
precipitation and can be distinguished from the initial rate-determining steps 
discussed here\cite{Luong2023_nanoscale}. 
Nevertheless, this subsequent diffusion-limited process is undoubtedly critical for 
achieving the complete conversion of the MgO solid.
%
%
\section{Conclusions} \label{sec:Conclusion}

We have elucidated molecular processes in MgO hydration by investigating a variety of 
interfacial structures obtained through a computational chemistry approach, 
including PS-MD simulations and DFT calculations.
In this study, we particularly focused on three fundamental processes of the hydration 
reaction: the dissociative adsorption of interfacial water, the dissolution of \mgion{}
ions at the early stage, and the nucleation and growth of \mgoh. 
Based on the findings of this study, 
the molecular process of the hydration reaction can be summarized as follows.

First, upon sufficient adsorption, water molecules strongly bound on the MgO surface 
form a geometrically distorted two-dimensional HB network in the first layer. 
This HB network triggers the dissociation of downward-oriented water molecules.
We have demonstrated that the next step following the dissociative water adsorption  
is the dissolution of \mgion{} ions from the surface.
This dissolution can proceed exothermically even from the defect-free surface
with an average activation barrier of $\sim$12 kcal/mol, indicating that 
\mgion{} dissolution is both thermodynamically and kinetically feasible. 
However, not all \mgion{} ions exothermically dissolve from the surface.
In terms of a molecular and electronic structure, the most important factor 
driving exothermic dissolution was found to be the sufficient stabilization of 
the resulting vacancy through proton adsorption onto neighboring oxygen atoms. 
Thus, the dissolution of \mgion{} ions is revealed to be a concerted process, 
involving not only the coordination by water molecules and hydroxide ions 
but also the water dissociation.
This result also indicates that the dissolution is a highly 
heterogeneous process that strongly depends on the local HB network structure.
Following the initial dissolution of \mgion{} ions, the hydration continues with 
the further dissolution of \mgion{} ions and the concurrent 
penetration of protons into the MgO solid. 
This proton penetration is crucial for compensating for the negative charge vacancy 
created by the dissolved \mgion{} ions. 
The dissolution of \mgion{} ions from an upper layer opens pathways, 
facilitating the subsequent \mgion{} dissolution from the sublayers. 
The fate of these dissolved \mgion{} ions is twofold: 
while some are incorporated into a newly formed MgO layer on the original MgO surface, 
others migrate to the outer water layer. 
This migration is thought to promote further \mgion{} dissolution from the crystal lattice. 
It is also expected that these newly exposed \mgion{} ions attract additional 
water molecules from the gas phase when the hydration is operated in a humid atmosphere.
Finally, \mgoh{} nucleation and growth proceeds according to the dissolution-precipitation 
mechanism rather than the solid-state reaction mechanism. 
Nucleation and growth via the solid-state reaction mechanism is difficult because 
it requires the removal of \mgion{} ions from the subsurface while maintaining 
the surrounding MgO (111) layer. 
This process and the relevant structural relaxation are severely hindered by the strong 
lattice potential from the MgO crystal environment. 
In contrast, the dissolution-precipitation mechanism involves \mgion{} ions that have moved 
far enough from the MgO surface to escape the influence of the lattice potential. 
These solvated ions then combine with OH$^-$ ions within the water layer to form chain-like 
structures, which act as precursors to the \mgoh{} crystal nuclei. 
This process is greatly facilitated by the presence of sufficient water, as evidenced by 
the spontaneous formation of a well-ordered crystalline \mgoh-like nucleus 
in calculations of a bulk water environment. 
This result demonstrates that an adequate aqueous environment is essential 
for efficient nucleation and growth.

Although the hydration of MgO is sensitive to the experimental conditions and is also
a non-equilibrium process probably without a single and unique reaction pathway, 
we have revealed a series of plausible, locally stable intermediate structures 
and the relevant molecular processes governing the chemical reaction on its complex 
energy landscape. 
We view this work as a foundational step toward computationally elucidating the mechanisms 
of this class of complex and multi-step solid-surface chemical reactions.


\begin{acknowledgement}
This work was supported by JSPS KAKEMHI Grant Numbers JP21K14723 and JP25K08562 to T.I. 
The calculations were partly carried out at the Academic Center for
Computing and Media Studies (ACCMS) at Kyoto University
and at Research Center for Computational Science, Okazaki,
Japan (Projects: 23-IMS-C058, 24-IMS-C057, and 25-IMS-C058).
\end{acknowledgement}

\begin{suppinfo}
 Details of potential-scaling molecular dynamics simulations for structure generation, 
 density functional based tight-binding and density functional theory calculations for 
 structure refinement, calculation system and conditions, nudged elastic band calculations
 for \mgion{} dissolution, and potential-scaling molecular dynamics simulations 
 for nucleation in water;
 Potential energy variation of interfacial structures;
 and Supplementary tables and figures noted in the main text.
 Parameter file for reactive force field (ReaxFF) calculations (ffield.txt).
\end{suppinfo}


\clearpage

\bibliography{references}

%
%
\begin{figure}
\begin{center}
\textbf{Table of Contents (TOC) graphics.}
 \includegraphics[width=11cm]{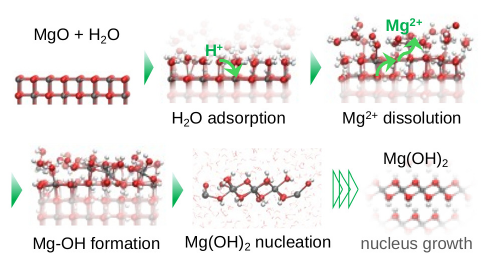}
\end{center}
\end{figure}


\end{document}